\documentclass[aps,prl,floatfix,twocolumn,superscriptaddress]{revtex4}
\usepackage{graphicx}
\usepackage{amssymb}
\usepackage{amsfonts}
\usepackage{textcomp}

\begin{document}

\title{High quality mechanical and optical properties of commercial silicon
nitride membranes}

\author{B. M. Zwickl}
\email[Electronic mail: ]{benjamin.zwickl@yale.edu}
\affiliation{Department of Physics, Yale University, 217 Prospect Street, New Haven, CT 06511}
\author{W. E. Shanks}
\affiliation{Department of Physics, Yale University, 217 Prospect Street, New Haven, CT 06511}
\author{A. M. Jayich}
\affiliation{Department of Physics, Yale University, 217 Prospect Street, New Haven, CT 06511}
\author{C. Yang}
\affiliation{Department of Physics, Yale University, 217 Prospect Street, New Haven, CT 06511}
\author{A. C. Bleszynski Jayich}
\affiliation{Department of Physics, Yale University, 217 Prospect Street, New Haven, CT 06511}
\author{J. D. Thompson}
\affiliation{Department of Physics, Yale University, 217 Prospect Street, New Haven, CT 06511}
\author{J. G. E. Harris}
\affiliation{Department of Physics, Yale University, 217 Prospect Street, New Haven, CT 06511}
\affiliation{Department of Applied Physics, Yale University, 15 Prospect Street, New Haven, CT 06511}

\date{\today}

\begin{abstract}
We have measured the optical and mechanical loss of commercial silicon nitride membranes.  We find that 50 nm-thick, 1 mm$^2$ membranes have mechanical $Q > 10^{6}$ at 293 K, and $Q > 10^7$ at 300 mK, well above what has been observed in devices with comparable dimensions.  The near-IR optical loss at 293 K is less than $2\times 10^{-4}$.  This combination of properties make these membranes  attractive candidates for studying quantum effects in optomechanical systems.
\end{abstract}


\maketitle

There is considerable interest in studying systems in which optical and mechanical degrees of freedom are coupled via radiation pressure.  These optomechanical systems offer a promising architecture for controlling the quantum states of light and matter, and for exploring the boundary between quantum and classical mechanics\cite{bose99}. Theoretical proposals suggest that optomechanical devices consisting of an optical cavity detuned by the motion of a micromechanical oscillator should be able to generate squeezed light \cite{hilico92}, entangled optical and mechanical degrees of freedom \cite{vitali07}, superposition states of mechanical resonators \cite{marshall03}, and QND measurements of a mechanical oscillator's energy \cite{thompson07}.  

The primary technical challenge to achieving these goals has been integrating high finesse $\mathcal{F}$ optical cavities with high quality factor $Q$ micromechanical oscillators. High optical $\mathcal{F}$ is desirable because it increases the optomechanical coupling per incident photon, and high mechanical $Q$ is desirable because it reduces the oscillator's coupling to the environment. Most experiments to date have used optomechanical systems in which an optical cavity is formed between a macroscopic mirror and a second mirror integrated into a micromechanical element. However it has been difficult to realize a high-$\mathcal{F}$ mirror and a high-$Q$ micromechanical resonator in a single device. An alternate approach is to place a thin dielectric slab between two macroscopic, rigid, highly reflective mirrors.  The slab acts as the mechanical oscillator while the macroscopic mirrors define the optical cavity. Optomechanical coupling arises because the cavity detuning depends upon the position of the dielectric slab \cite{thompson07}.

For this approach to work the dielectric slab must have high $Q$ and not substantially decrease the cavity $\mathcal{F}$. In this paper we describe commerically available silicon nitride membranes which meet these criteria. These membranes should make it feasible to integrate $Q = 10^7$ mechanical oscillators with finesse $\mathcal{F} \gtrsim 3\times 10^5$ cavities, pushing optomechanical systems substantially closer to the quantum regime than has been possible previously.  


Commercially available low-stress SiN$_{x}$ membranes are used as vacuum windows for x-ray spectroscopy and as sample holders for transmission electron microscopy (TEM).  They have remarkable strength against static loads given their dimensions, as well as high surface smoothness and flatness, chemical inertness, and transparency to visible light, x-rays, and electrons \cite{toivola03}. However, little was known about their dynamic mechanical properties or near-IR optical properties (other than the real part of the index of refraction, $n$).  In this paper we present measurements of these properties for membranes from two manufacturers, and discuss their figures of merit for various applications.

\begin{figure}
\includegraphics[width=0.35\textwidth]{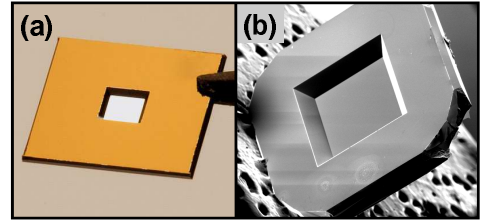}%
\caption{\label{fig:schematic} (a) Photo of a 1 mm$\times$1 mm$\times$50 nm Norcada x-ray membrane.  (b) SEM of an SPI 1 mm$\times$1 mm$\times$50 nm membrane.}
\end{figure}

Three models of low-stress SiN$_{x}$ membranes from Norcada Inc.\ \cite{norcada} and Structure Probe Inc.\ (SPI) \cite{spi} were studied.  All three models were 1 mm square membranes suspended from a 200 {\textmu}m thick Si frame.  The Norcada models were 50 nm and 100 nm-thick x-ray windows (Fig.\ \ref{fig:schematic}(a)), while the SPI membrane was a 50 nm-thick TEM window (Fig.\ \ref{fig:schematic}(b)).  

For room temperature mechanical measurements, the membrane is placed at the center of a  $\mathcal{F}\sim 1$ Fabry-Perot cavity.  A 1550 nm diode laser monitors the position of the membrane interferometrically.  The membrane is mounted on a piezo-electric actuator.  The cavity and membrane are mounted in a vacuum chamber evacuated by an ion pump to a pressure $\leq 10^{-6}$ Torr.

\begin{table}
\caption{\label{tab:Qtable}Table of measured resonant frequency and $Q$ for several vibrational modes of a Norcada 50 nm-thick membrane at $T = 293$ K.  For all modes except (1,1) the deviation from the theoretical resonant frequency $\Delta\nu_{i,j} = \nu_{i,j}-\nu_{1,1}\sqrt{(i^2+j^2)/2}$ is given.}
\begin{ruledtabular}
\begin{tabular}{lcccc}
Mode & i = 1 & 2 & 3 & 4 \\
\hline
j = 1 & $\nu_{1,1} = $ 133,784 & $\Delta\nu_{1,2} = 168$ Hz & 301 Hz & 247 Hz \\
 & $Q = 1.1\times 10^6$ & $1.1\times 10^6$ & $2.2\times 10^6$ & $1.5\times 10^6$ \\
\hline
2 & -82 Hz & 154 Hz & 380 Hz & --- \\
 & $1.6\times 10^6$ & $0.8\times 10^6$ & $1.8\times 10^6$ & --- \\
\hline
3 & ---  & 17 Hz & 177 Hz & ---\\
 & ---  & $1.8\times 10^6$ & $1.9\times 10^6$ & --- \\
\hline
4 & -244 Hz  & ---  &  --- & ---  \\
  & $1.5\times 10^6$  & ---  & ---  & --- \\
\end{tabular}
\end{ruledtabular}
\end{table}

\begin{figure}
\includegraphics[width=.45\textwidth]{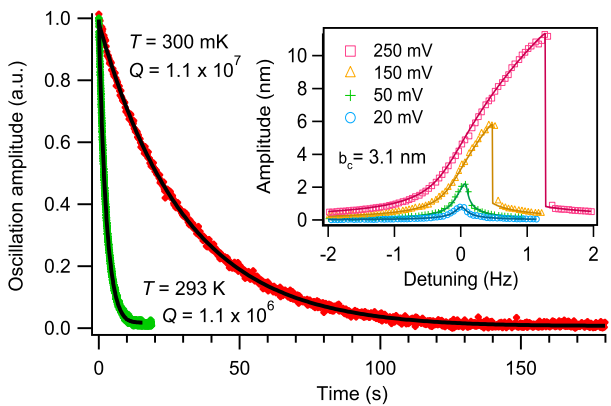}%
\caption{\label{fig:ringdown}Mechanical ringdown of the membrane.  The membrane was driven by a piezo actuator which was stopped at $t = 0$.  Inset shows the non-linear mechanical response of the Norcada 50 nm membrane.  Solid lines are fits.}
\end{figure}

Table \ref{tab:Qtable} shows the room temperature resonant frequency and $Q_{i,j}$ for several vibrational modes of a Norcada 50 nm-thick membrane, where $i$ and $j$ are positive integers indexing the membrane's vibrational modes. The eigenfrequencies for a stressed square membrane are $\nu_{i,j} = \nu_{1,1} \sqrt{(i^2 + j^2)/2}$. The resonant frequencies in Table \ref{tab:Qtable} agree with this expression to within $1\times 10^{-3}$.  Each mode shown in Table \ref{tab:Qtable} has a $Q > 8\times 10^5$. Each $Q_{i,j}$ was determined by using the piezo to drive the membrane at $\nu_{i,j}$, then switching off the drive and monitoring the mechanical ringdown. Fig.\ \ref{fig:ringdown} shows one such measurement for the fundamental ($i$ = 1, $j$ = 1) vibrational mode of the 50 nm Norcada membrane, which has $Q = 1.1\times 10^6$.  The spring constant for this mode is $k_{1,1} = 30$ N/m.  The mechanical behavior of the 50 nm-thick SPI and 100 nm-thick Norcada membranes is similar to the data in Table \ref{tab:Qtable} but with $Q_{i,j} \sim 10^5$.  

Recent theoretical work has shown that quantum behavior may be observable in the response of a nonlinear mechanical oscillator \cite{katz07}.  The inset in Fig.\ \ref{fig:ringdown} shows the nonlinear response of a 50 nm-thick Norcada membrane at room temperature.  Four different drive amplitudes are shown.  The data is fit to the response of a Duffing oscillator \cite{nayfeh79} with equation of motion $\ddot{z} + (\omega_{0}/Q )\dot{z} + {\omega_{0}}^2 z + \beta z^3 = F_{0} \sin(\omega t)$.  From the fits we find $\beta = (4/3)^{3/2}{\omega_0}^2/Q {b_{\text{c}}}^2 =  1.3\times 10^{23}$ m$^{-2}$s$^{-2}$, which is too small for these membranes to show the quantum effects discussed in \cite{katz07}.

\begin{figure}
\includegraphics[width=0.4\textwidth]{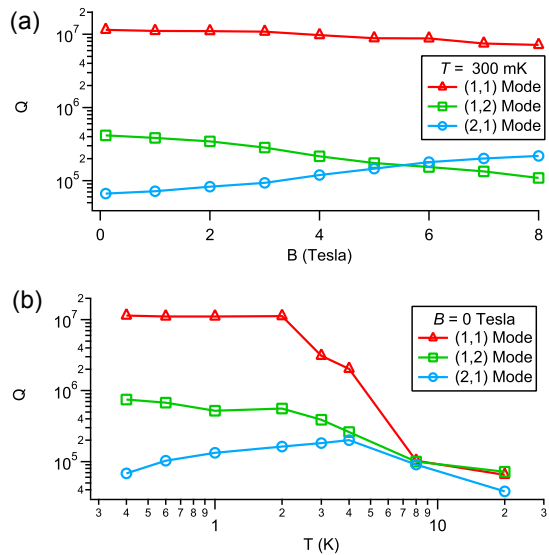}%
\caption{\label{fig:QvsBandT} (a) Comparison of $Q$ versus magnetic field ($T = 300$ mK) for the three lowest vibrational modes of the 50 nm Norcada membrane.  (b) Comparison of $Q$ versus temperature ($B = 0$) for the three lowest vibrational modes of the 50 nm Norcada membrane.}
\end{figure}

The Norcada membranes were also studied extensively at cryogenic temperatures in a $^{3}$He refrigerator.  For these measurements, the membrane position was monitored using a fiber optic interferometer \cite{rugar89}.  A piezo-electric actuator is used to drive the membrane.  The fridge is surrounded by a superconducting solenoid.     

Fig. \ref{fig:QvsBandT}(a) shows $Q$ as a function of $B$ (applied normal to membrane) at $T = 300$ mK.  Although $Q$ varies slightly with $B$, $Q_{1,1} \approx 10^7$ over this range of magnetic fields.  Fig.\ \ref{fig:ringdown} shows a typical mechanical ringdown measurement at $T = 300$ mK giving $Q = 1.1\times 10^7$.  

The temperature dependence of $Q$ for the three lowest vibrational modes is shown in Fig.\ \ref{fig:QvsBandT}(b).  The fundamental mode only achieves its maximum $Q_{1,1}$ for $T \lesssim$ 2 K, while the other  modes do not exceed $10^6$ at cryogenic temperatures.  Less comprehensive cryogenic measurements of SPI 50 nm-thick membranes revealed that their $Q_{i,j}$ never exceeds $10^6$ even at $T = 300$ mK.  It is not clear what mechanism limits $Q$, but we note that the limit set by thermoelastic dissipation is $3\times 10^{11}$ at $T = 293$ K and should only increase at cryogenic temperatures \cite{zener37,norris05}.

\begin{figure}
\includegraphics[width=0.4\textwidth]{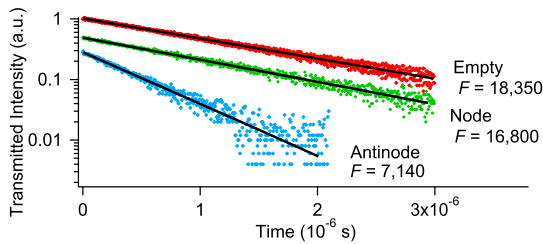}%
\caption{\label{fig:finesse} Ringdown of transmission from cavity after the laser is chopped at $t = 0$, shown with fitted values of finesse. }
\end{figure}

Finally, we measured the $\lambda = 1064$ nm optical properties of these membranes at room temperature using a $\mathcal{F} \simeq 18000$ cavity.  Placing a membrane inside the optical cavity can lead to diminished cavity $\mathcal{F}$, either from absorption or diffuse scattering by the membrane. In order to determine the optical loss, we first measured $\mathcal{F}$ without a membrane in the cavity, and then with the membrane inserted at a node and antinode of the cavity optical field. At each membrane position $\mathcal{F}$ was determined by ringdown measurements of the $\lambda = 1064$ nm intracavity field \cite{rempe92}. 

Fig. \ref{fig:finesse} shows the optical power transmitted through the cavity as a function of time $t$ after the incident beam is switched off by an AOM at $t$ = 0.  By fitting the full dependence of $\mathcal{F}$ with membrane position we determine the imaginary part of the index of refraction $k_{1064} = 1.6 \pm 0.1 \times 10^{-4}$, giving a maximum round-trip cavity loss of $4\pi k l/\lambda = 2\times 10^{-4}$, where $l$ is twice the membrane thickness.  This value of $k$ is consistent with previously published results that indicate $k_{1064} \leq 1.0 \times 10^{-3}$ \cite{philipp73,poenar97}.  We observed similar loss in the other SPI and Norcada membranes.

Because the optical loss seems well-modeled by membrane absorption alone, we believe scattering at the membrane surface is negligible.  Low scattering loss is reasonable, as these membrane are esteemed for their smoothness, flatness, and cleanliness.  SPI estimates a surface roughness for their membranes $\lesssim$ 0.2 nm and surface flatness $\sim 1$ nm over the entire membrane, figures which are sufficient for super-polished high-$\mathcal{F}$ cavity mirrors \cite{rempe92}. 

Although there is no single figure of merit for the performance of a micromechanical device, it is useful to compare these membranes with other micromechanical devices using two parameters relevant to a broad range of applications. The first is the membrane's thermal force noise, $S_{\text{F}} = \sqrt{4 k k_B T/\omega_0 Q}$.  For the Norcada membrane at 300 mK, $S_{\text{F}} = 8\times 10^{-18}$ NHz$^{-1/2}$, within an order of magnitude of $S_{\text{F}}$ for the single-crystal silicon cantilever used to detect spin resonance of a single electron \cite{rugar2004,stowe97}. 

A second useful figure of merit is the $Q$ relative to the size (either thickness or volume).  There is a widely noted trend that smaller resonators have lower $Q$'s \cite{ekinci05,liu05} (for instance the 60 nm thick Si cantilevers in \cite{stowe97} have $Q = 6700$).  However, $Q = 1.1\times 10^7$ observed here for 50 nm thick Norcada membranes breaks sharply from this trend.  Such a large $Q$ is typically seen in the bulk vibrations of cm-scale single crystal silicon \cite{kleiman87}.  High quality factors in silicon nitride micro- and nano-resonators have been previously reported at cryogenic temperatures \cite{naik06}, and room temperatures \cite{verbridge06}, but these have not exceeded $2.1\times 10^5$.

In conclusion, silicon nitride membranes offer an outstanding combination of high force sensitivity and a large surface area in a commercially available device.  They have remarkably high $Q$ factors at both room and cryogenic temperatures and in the presence of large magnetic fields.  These properties combined with their small near-IR optical loss make them particularly well-suited for experiments involving radiation pressure in optical cavities, which typically require sensitive micromechanical devices with transverse dimensions large enough to accommodate an optical spot of 10--100 {\textmu}m.  These properties make silicon nitride membranes attractive in a broad range of applications of sensitive force detectors \cite{kenny01,waggoner07,lamoreaux97}.

This work was supported by  NSF grants 0555824 and 0653377, and the Yale Institute for Nanoscience and Quantum Engineering. B. M. Z. acknowledges the support of an NSF Graduate Research Fellowship, and J. G. E. H. acknowledges support from an Alfred P. Sloan Foundation fellowship.

\end{document}